\documentclass[]{spie}  

\usepackage{amsmath,amsfonts,amssymb}
\usepackage{graphicx}
\usepackage[colorlinks=true, allcolors=blue]{hyperref}
\usepackage{xcolor}
\usepackage[normalem]{ulem}
\usepackage{verbatim}
\usepackage{mwe}
\usepackage{subfig}
\usepackage{pgfplotstable,filecontents}
\pgfplotsset{compat=1.9}

\title{CCAT: Detector Noise Limited Performance of the RFSoC-based Readout Electronics for mm/sub-mm/far-IR KIDs}

\author[a]{Adrian K. Sinclair}
\author[a]{James Burgoyne}
\author[d]{Anthony I. Huber}
\author[f]{Colin Murphy}
\author[l]{Steve K. Choi}
\author[f]{Cody J. Duell}
\author[f]{Zachary B. Huber}
\author[f]{Yaqiong Li}
\author[a,d,e]{Scott C. Chapman}
\author[f]{Michael D. Niemack}
\author[i]{Thomas Nikola}
\author[f]{Eve M. Vavagiakis}
\author[f,p]{Samantha Walker}
\author[g]{Jordan D. Wheeler}
\author[g]{Jason Austermann}
\author[f]{Lawrence Lin}
\author[k]{Ruixuan Xie}
\author[o]{Bugao Zou}
\author[m]{Philip D. Mauskopf}

\author{the CCAT collaboration}

\affil[a]{Dept. of Physics and Astronomy, University of British Columbia, Vancouver, BC, Canada} 
\affil[d]{Dept. of Physics and Astronomy, University of Victoria, Victoria, BC, Canada }
\affil[e]{Herzberg Astronomy and Astrophysics Research Centre, National Research Council Canada, Victoria, BC, Canada}
\affil[f]{Dept. of Physics, Cornell University, Ithaca, NY, USA }
\affil[g]{National Institute of Standards and Technology, Boulder, CO, USA }
\affil[i]{Cornell Center for Astrophysics and Planetary Sciences, Cornell University, Ithaca, NY, USA}
\affil[j]{Dept. of Astronomy, Cornell University, Ithaca, NY, USA}  
\affil[k]{Dept. of Electrical Engineering, University of British Columbia, Vancouver, BC, Canada}
\affil[l]{Dept. of Physics and Astronomy, University of California, Riverside, CA, USA}
\affil[o]{Dept. of Applied and Engineering Physics, Cornell University, Ithaca, NY, USA}
\affil[m]{School of Earth and Space Exploration and Department of Physics, Arizona State University, Tempe, AZ, USA}
\affil[p]{Cornell Center for Materials Research, Cornell University, Ithaca, NY, USA}

\authorinfo{Further author information: (Send correspondence to A.K.S.)\\A.K.S.: E-mail: adriansinclair@phas.ubc.ca}

\begin{document} 
\maketitle

\begin{abstract}
The Fred Young Submillimeter Telescope (FYST), on Cerro Chajnantor in the Atacama desert of Chile, will conduct wide-field and small deep-field surveys of the sky with more than 100,000 detectors on the Prime-Cam instrument. Kinetic inductance detectors (KIDs) were chosen as the primary sensor technology for their high density focal plane packing. Additionally, they benefit from low cost, ease of fabrication, and simplified cryogenic readout, which are all beneficial for successful deployment at scale. The cryogenic multiplexing complexity is pulled out of the cryostat and is instead pushed into the digital signal processing of the room temperature electronics. Using the Xilinx Radio Frequency System on a Chip (RFSoC), a highly multiplexed KID readout was developed for the first light Prime-Cam and commissioning Mod-Cam instruments. We report on the performance of the RFSoC-based readout with multiple detector arrays in various cryogenic setups. Specifically we demonstrate detector noise limited performance of the RFSoC-based readout under the expected optical loading conditions. 


\end{abstract}

\keywords{CCAT, RFSoC, MKID, frequency multiplexed, kinetic inductance}

\section{Introduction}

The Fred Young Submillimeter Telescope (FYST)\cite{Parshley2018}, which will be located on Cerro Chajnantor in the Atacama desert, will host the Mod-Cam\cite{Vavagiakis2018}, then Prime-Cam instrument, which will be capable of supporting more than 100,000 kinetic inductance detectors (KIDs). The KID arrays are being developed by the Quantum Sensors Division at NIST. The broadband imaging polarimeter channels are being designed for background limited operation at the high altitude, 5600-meter FYST site. Background limited operation implies that the noise of the system is limited by the statistics of the incoming photons. 
Ground-based observatories are typically dominated by atmospheric photons with contributions from the warm optics and the astronomical sky signal. Maintaining this background limited operation sets up a noise hierarchy for the rest of the detector biasing and readout electronics i.e., $T_{\text{photons}} > T_{\text{detector}} > T_{\text{readout}}$ would be the preferred scaling of noise temperatures. The two main readout components for Mod-Cam and Prime-Cam are the cryogenic low noise amplifiers and the RFSoC-based frequency multiplexed readout systems \cite{Sinclair2022}. The cryogenic amplifiers have well characterised noise and their performance is mostly independent of the detectors. The RFSoC-based readout however, which provides the active biasing and sensing for the detectors, interlinks the readout and detector properties.    

In this proceeding, we present the setup and measurements of the RFSoC-based readout system on multiple KID arrays, the results of which show that detector noise is greater than readout noise in certain cases relevant for operation. We then discuss the detector parameters that lead to the desired noise hierarchy and have the potential to significantly relax readout noise requirements.   

\section{Readout and Measurement Setup}
\begin{figure}
\centering
\includegraphics[width=0.35\linewidth]{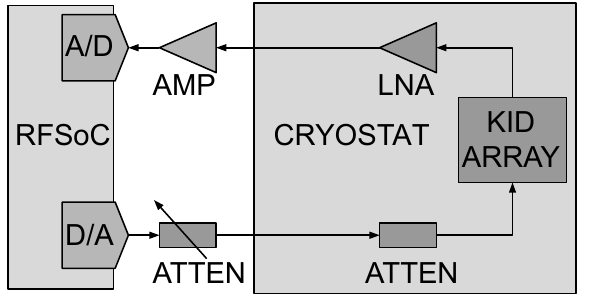}
\caption{General measurement setup with the RFSoC-based readout system on the left connected to the cryostat and detectors (KID) on the right. Starting from the digital-to-analog converters (D/A) and following the signal path, the generated bias tones enter a programmable attenuator at room temperature then a few stages of cryogenic attenuators before the detectors. After bias tones are modulated by the detectors they are amplified by a cryogenic low noise amplifier then amplified again at room temperature before being sampled by the RFSoCs analog-to-digital converters (A/D).}
\label{fig:setup}
\end{figure}
A block diagram of the measurement setup is shown in figure \ref{fig:setup}, with the relevant components labeled. 
Two cryostats were used in this analysis, one at Cornell University and the other at NRC Herzberg, both utilizing BlueFors dilution refrigerators to cool the detectors to milli-Kelvin temperatures. The detector arrays are situated on the mixing chamber plate (coldest plate) and connected via SMA coaxial cables to the outside of the cryostat. Along the path on the driving signal side are cryogenic capable attenuators at various stages to minimize the thermal noise propagation to the detectors and roughly set the power level for the biasing frequency comb. On the sensing, or outward path, from the detectors the modulated frequency comb is amplified by a cryogenic low noise amplifier on the 4 Kelvin stage of the cryostat. At room temperature the readout electronics consist of attenuators, a sensing side amplifier, and an RFSoC ZCU111 development board.

The RFSoC-based readout electronics use the ZCU111 development board and run custom FPGA gateware, ARM firmware, and software. For more information about the readout see the proceedings \cite{Sinclair2022, James2024}. For all measurements the RFSoC was running the \verb|primecam_readout| control software\cite{primecamreadout-github} and the current version of the digital design\cite{gateware-github}.

\section{Results}
\subsection{280 GHz KID Array}
The first array is an Aluminum KID witness chip for one array of the 280 GHz broadband camera module of the Mod-Cam (and later Prime-Cam) instrument. The witness chip had 5 active detectors as opposed to the science grade arrays which will have 3450 per array. A cryogenic blackbody source \cite{choi2018} was positioned in the detector array's optical path, along with aluminum feedhorns and filterstack. The combination of the filter and feedhorn waveguide defined a band that was centered at approximately 270 GHz and was nearly 20 GHz wide with about 60$\%$ optical efficiency. Different optical loads were presented to the detectors by varying the temperature of the thermal blackbody source.
Four measurements were analyses in this setup for this proceeding: sweeps and timestreams, both on and off-resonance, with the thermal load at 2.9 K and 13.2 K. The higher end of the thermal load (13.2 K), in combination with the filter stack, delivers an optical power that is comparable to the expected on-sky loading for FYST at 280 GHz \cite{choi2020, CCATSciForcast2021}

The on-resonance measurement will give the sum of all noise sources and the off-resonance will contain all readout noise but minimal noise from the detector. The difference between the noise observed on and off should be the most direct way of determining detector noise limited performance for a system. This assumes that the off-resonance measurement does not increase the noise from power saturation effects in the amplifiers or data-converters. It is also assumed the readout noise is approximately constant for the relatively small shift in frequency compared to the entire readout bandwidth for an off-resonance measurement.
We use the following equation to determine how far away from resonance the tone should be moved to ensure minimal contribution from the detector in an off-resonance measurement,
\begin{equation}
T(x)/T(0) = \frac{1}{(1 + 4 Q_r^2 x^2)^2} = \frac{1}{(1 + 4 N^2)^2}.
\label{eq:norm_off_noise}
\end{equation}
Where $T(x)/T(0)$ is the detector noise temperature\cite{Sinclair2022} normalized by its value on-resonance. $x = f/f_0 - 1$ is the fractional frequency shift from resonance, $Q_r$ is the resonator total quality factor, and $N = x Q_r$ is the number of linewidths. Using this equation we find that the detector noise drops to a hundredth of its value on resonance at 1.5 linewidths away. Off-resonance noise was measured 1 MHz off which is greater than 5 linewidths (-40 dB) away for all five detectors therefore we do not expect any appreciable detector noise contribution.

\begin{figure}
\includegraphics[width=.33\linewidth]{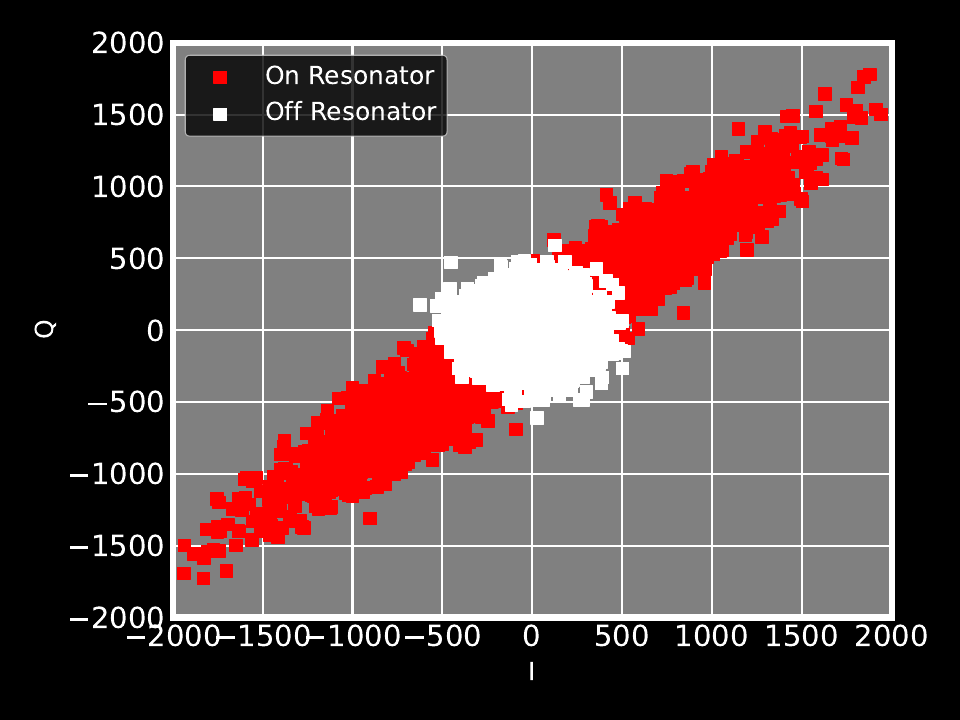}\hfill
\includegraphics[width=.33\linewidth]{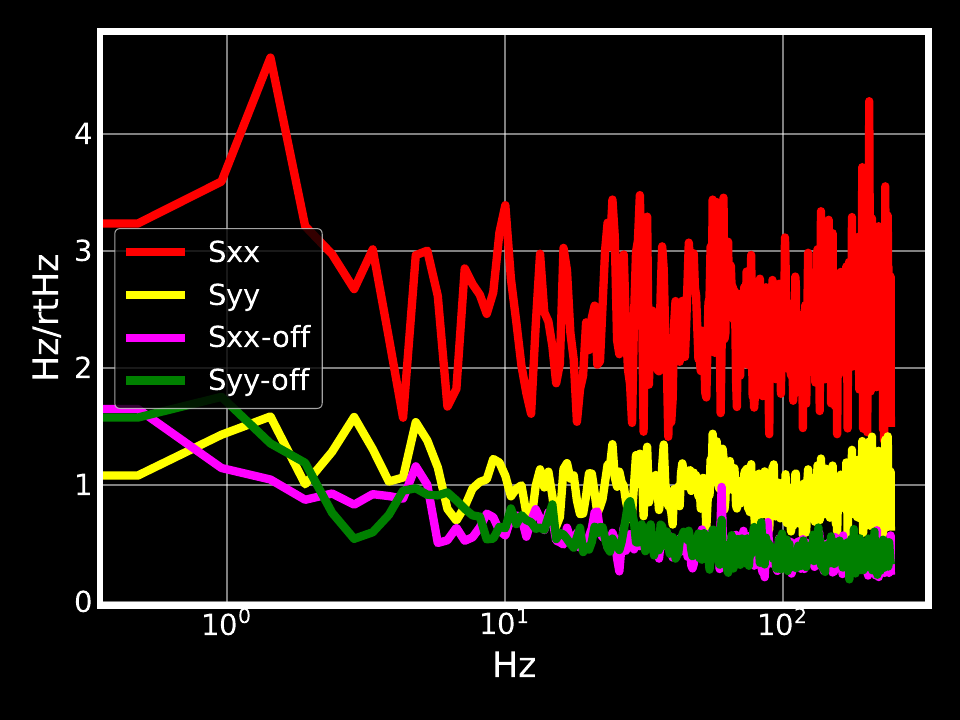}\hfill
\includegraphics[width=.33\linewidth]{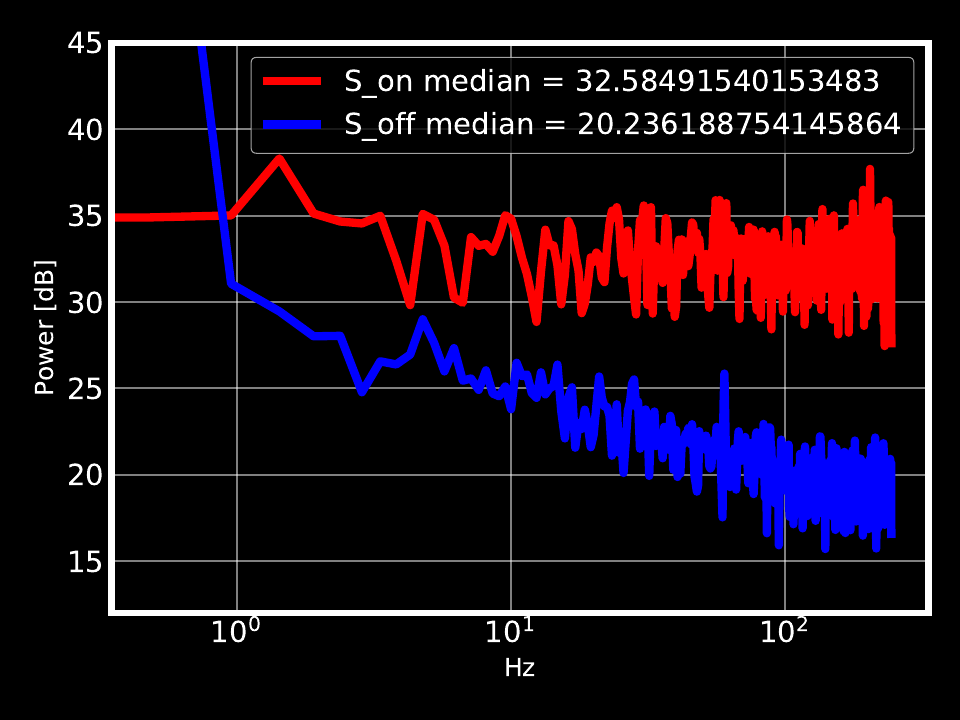}
\newline
\includegraphics[width=.33\linewidth]{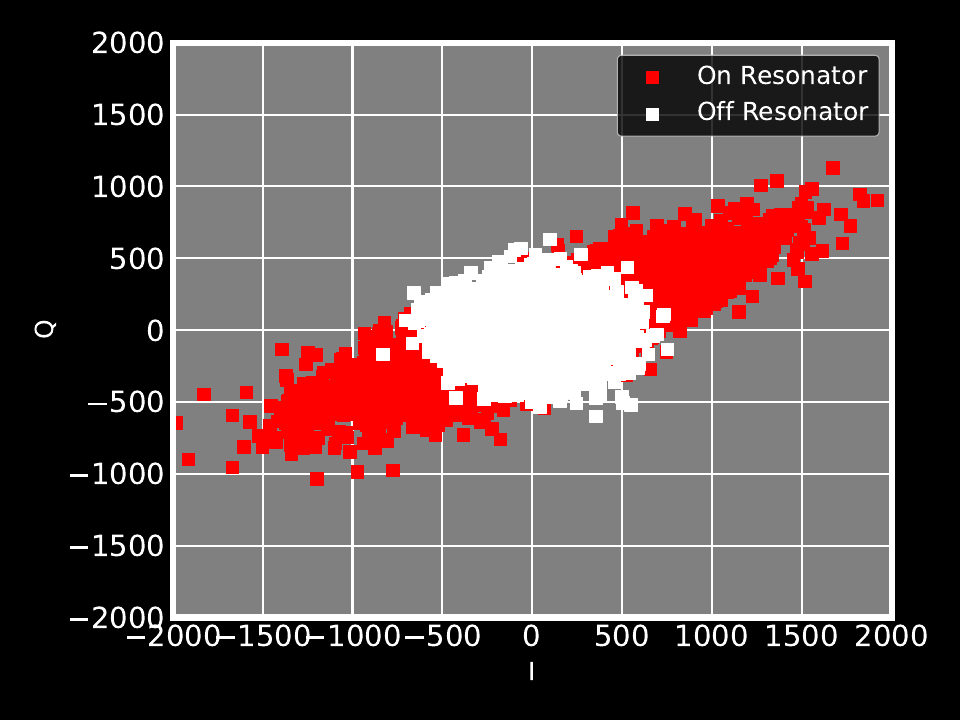}\hfill
\includegraphics[width=.33\linewidth]{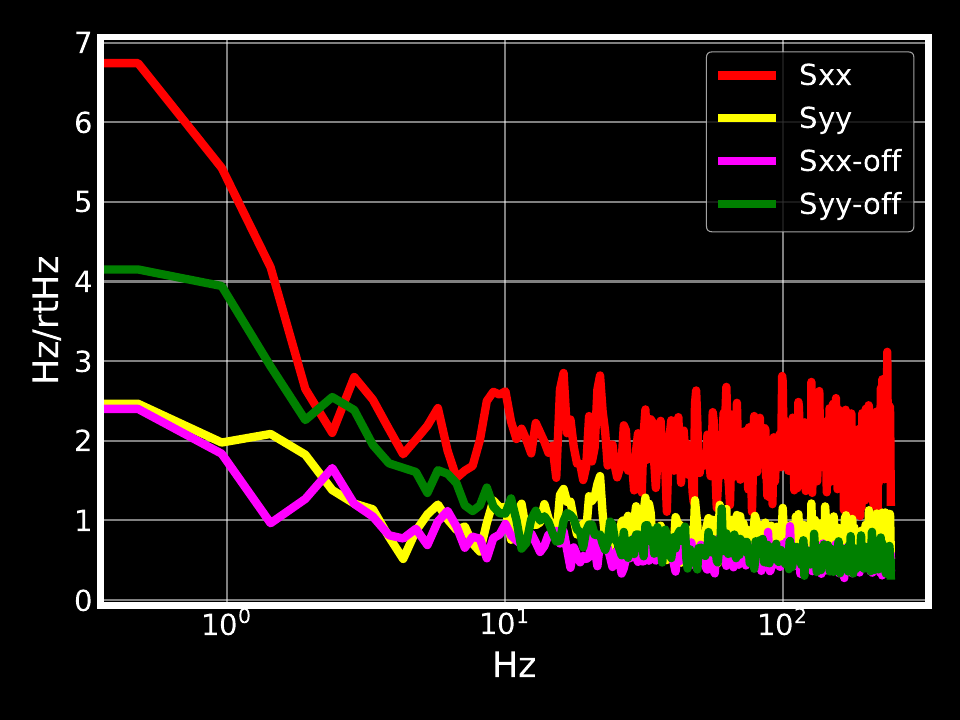}\hfill
\includegraphics[width=.33\linewidth]{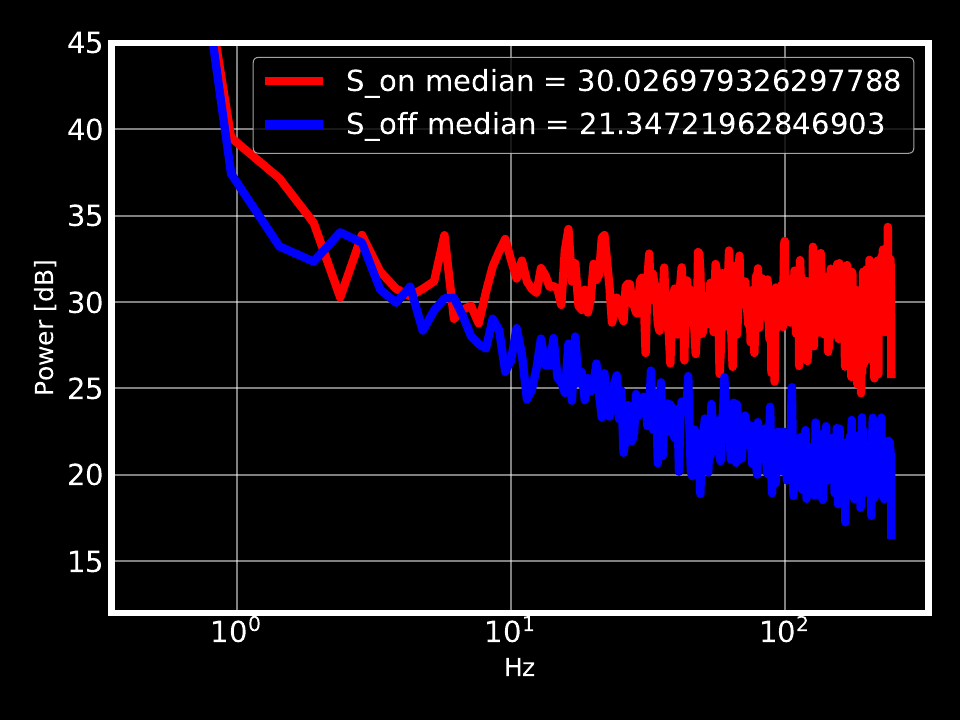}\hfill
\newline
\includegraphics[width=.33\linewidth]{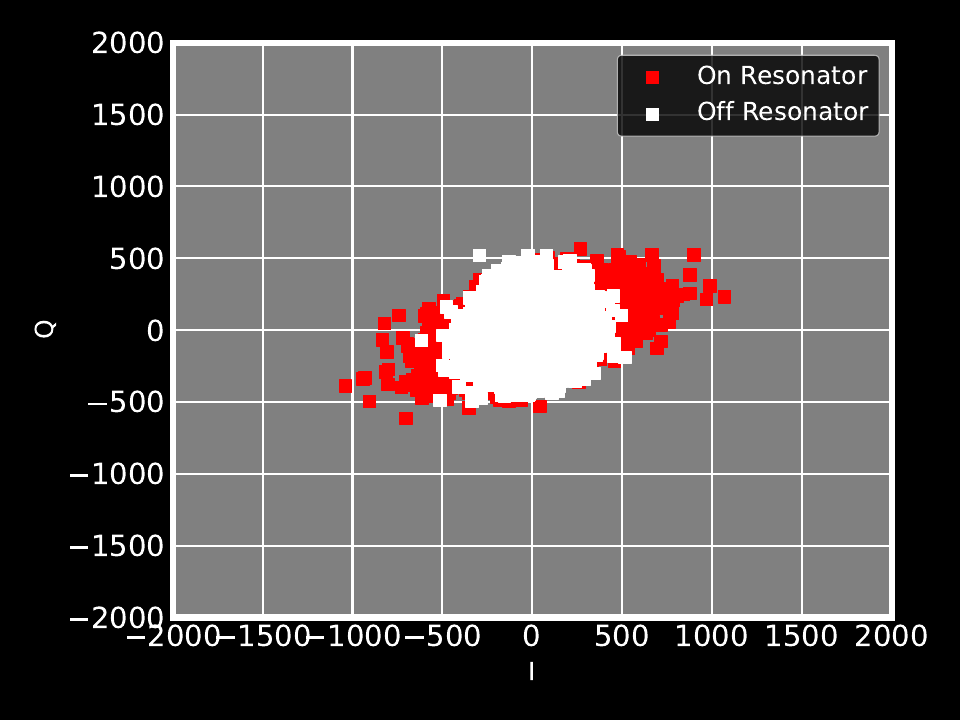}\hfill
\includegraphics[width=.33\linewidth]{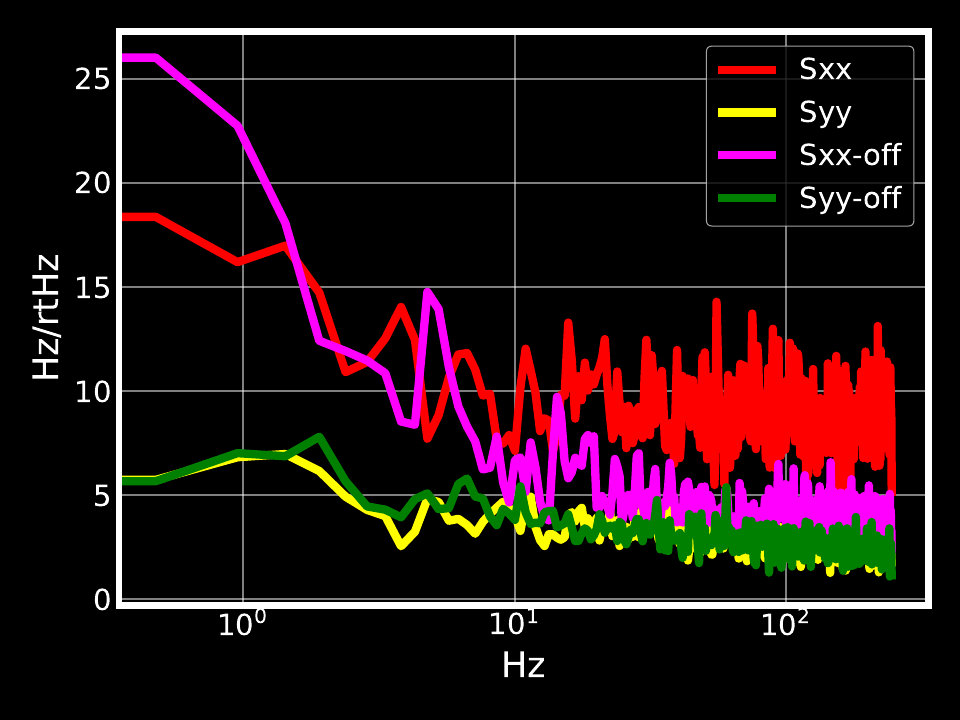}\hfill
\includegraphics[width=.33\linewidth]{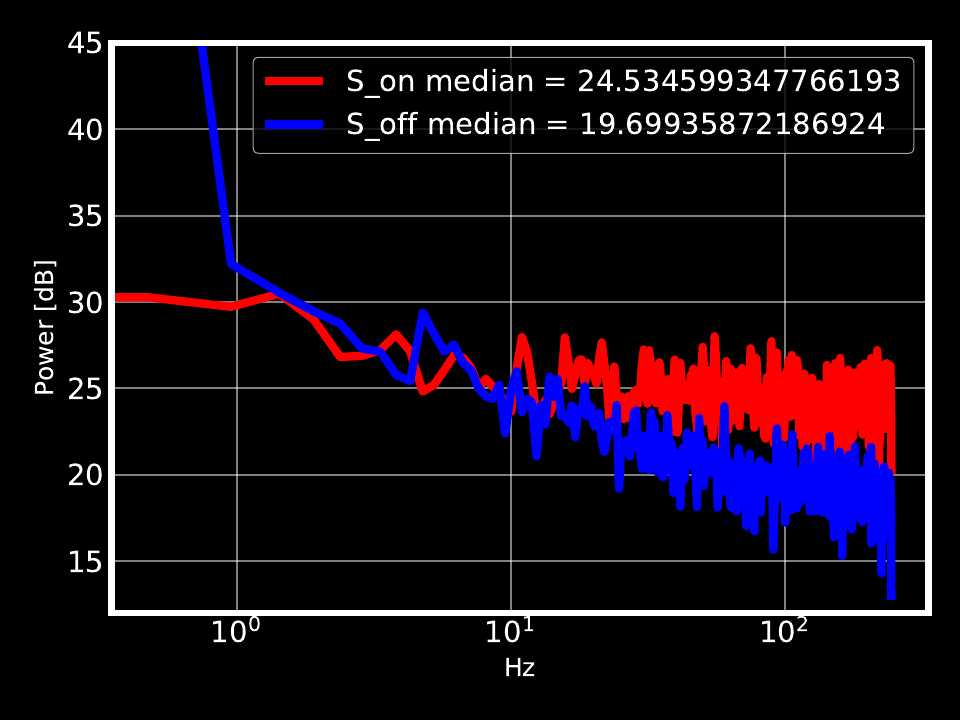}\hfill
\newline
\includegraphics[width=.33\linewidth]{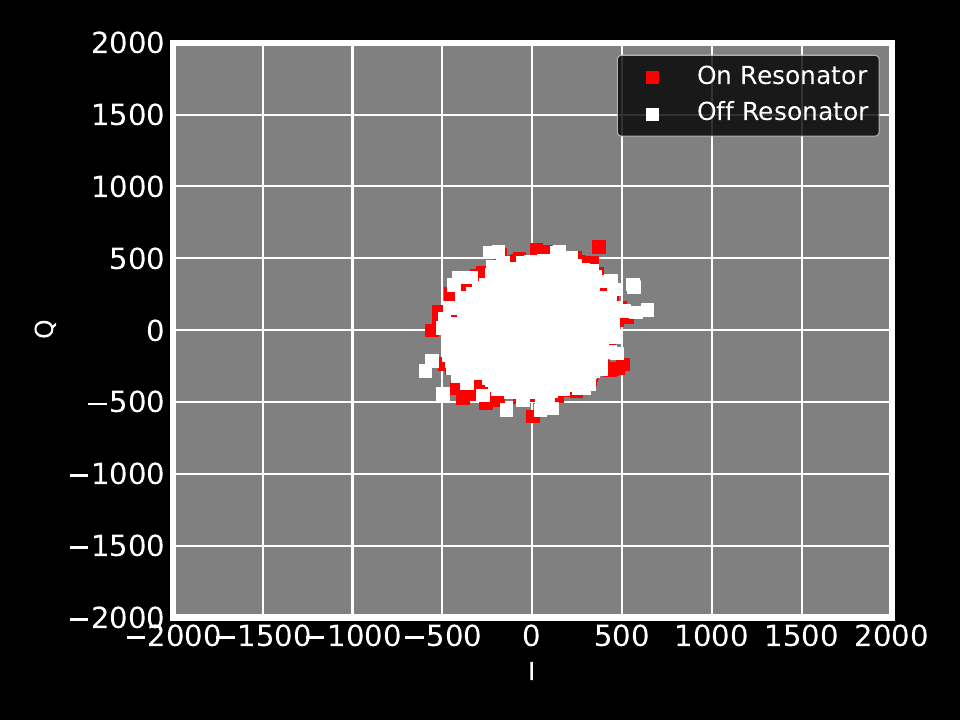}\hfill
\includegraphics[width=.33\linewidth]{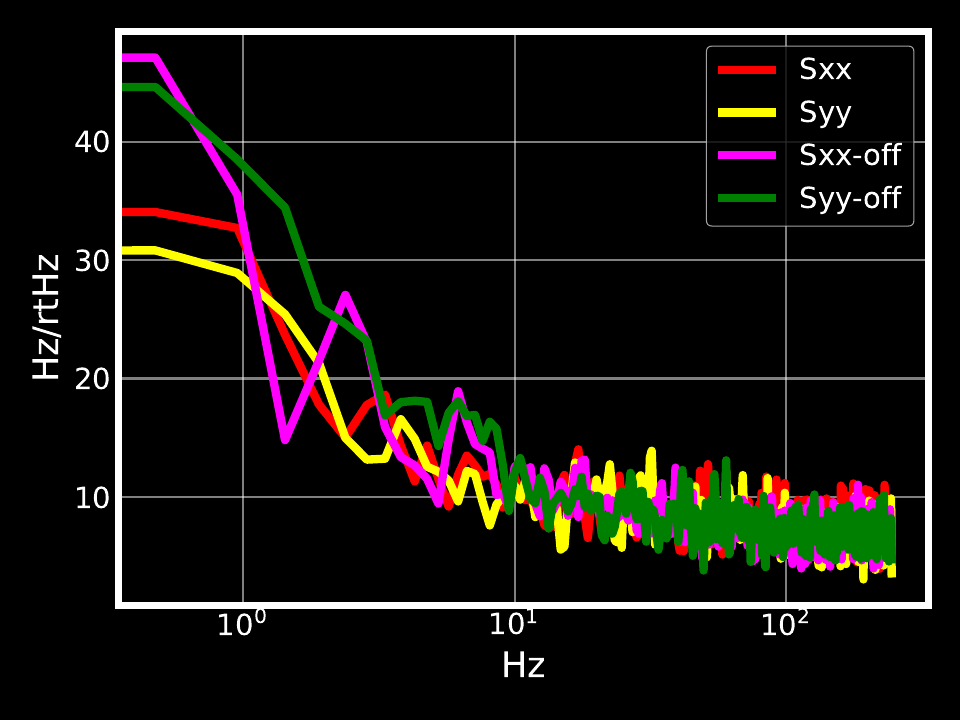}\hfill
\includegraphics[width=.33\linewidth]{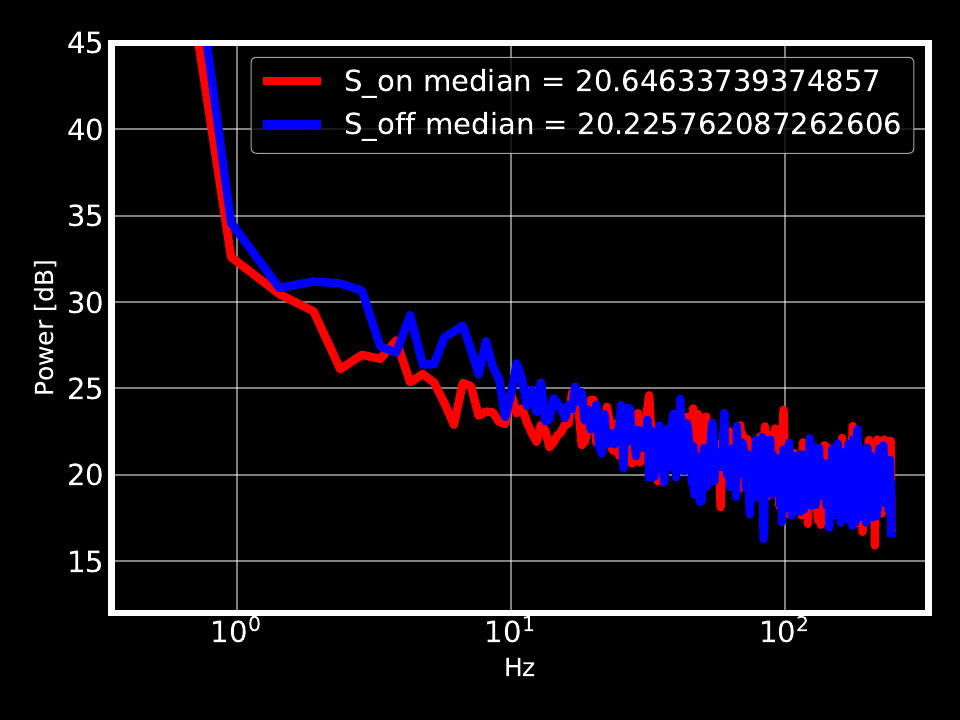}\hfill
\caption{First column shows the IQ plane scatter plot of the timestreams, the second column is the noise power spectral density in the parallel (xx) and perpendicular (yy) frequency quadratures, the third column gives the noise power spectral density of the quadrature sum of the noise in I and Q i.e. $\sqrt{S_{II}^2 + S_{QQ}^2}$. The first row is detector 1 under 2.9 K thermal loading, the second row is detector 1 under 13.2 K thermal loading, the third row is detector 5 under 2.9 K thermal loading, the fourth row is detector 5 under 13.2 K thermal loading.
}
\label{fig:280ghz}
\end{figure}

A more detailed look at the results of the RFSoC measurements with the 280 GHz witness array are given in figure \ref{fig:280ghz}.
Two out of the five detectors measured are highlighted for their differences in noise properties. A high quality factor resonator in the top two rows, and a low quality factor resonator in the bottom two rows. The two rows for each detector represent two different thermal load temperatures of 2.9 K and 13.2 K respectively. The first column is the IQ plane representation of the measured timestreams with medians subtracted for both on and off-resonance. Clearly the noise ellipse is larger on-resonance than off and higher loading is seen to reduce the on-resonance noise. Note that the tone powers were not modified between different loads. The second column is the noise power spectral density of the frequency ($S_{xx}$) and dissipation ($S_{yy}$) quadratures (or parallel and perpendicular). This was calculated via the methods found in references\cite{Barry2014, Gordon2016, Gao2008, James2024}. The third column is the quadrature sum of the individual noise power spectral densities of the I and Q timestreams. It can be seen that at the high-frequency end of the power spectrum, the difference in on and off-resonance noise power is more than 10 dB for the high quality factor resonator and more than 3 dB for the low quality factor resonator under the 2.9 K loading. The higher loading reduces this difference in both resonators. 

The measurements for all five detectors are summarized in the table \ref{tab:params} with resonant frequencies and quality factors found with the Python package \texttt{scraps} \cite{Carter2016}. The measured off-resonance tone powers are also listed.

\begin{table}[]

\begin{tabular}{llllllll}
Detector - load & $f_0$ (Hz)   & $Q_c$       & $Q_i$       & $Q_r$      & $P_{\text{off}}$ (dB) & $S_{\text{on}}$ (dB) & $S_{\text{off}}$ (dB) \\
\hline
1 – 2.9 K       & 510537355 & 2.23E+04 & 3.12E+04 & 1.30E+04 & 114.41    & 32.59      & 20.24       \\
2 – 2.9 K       & 619162202 & 1.72E+04 & 2.85E+04 & 1.07E+04 & 113.4     & 30.6       & 20.67       \\
3 – 2.9 K       & 705161100 & 4.77E+04 & 1.64E+04 & 1.22E+04 & 113.65    & 30.65      & 20.9        \\
4 – 2.9 K       & 800041169 & 4.15E+04 & 1.56E+04 & 1.13E+04 & 112.31    & 24.49      & 19.6        \\
5 – 2.9 K       & 900975202 & 2.96E+04 & 1.63E+04 & 1.05E+04 & 111.6     & 24.53      & 19.7        \\
1 – 13.2 K      & 510321885 & 2.22E+04 & 1.59E+04 & 9.26E+03 & 114.37    & 30.02      & 21.35       \\
2 – 13.2 K      & 618900959 & 1.87E+04 & 1.28E+04 & 7.60E+03 & 113.29    & 26.15      & 20.37       \\
3 – 13.2 K      & 704847045 & 4.94E+04 & 7.93E+03 & 6.83E+03 & 113.54    & 22.22      & 20.99       \\
4 – 13.2 K      & 799689798 & 4.21E+04 & 7.31E+03 & 6.23E+03 & 112.3     & 20.8       & 20.56       \\
5 – 13.2 K      & 900604601 & 3.12E+04 & 7.79E+03 & 6.23E+03 & 111.6     & 20.65      & 20.23      
\end{tabular}
\caption{280 GHz KID array measurements and fit parameters. Columns from left to right are: Detector number and thermal load temperature, $f_0$ resonant frequency in Hz; $Q_c$ coupling quality factor; $Q_i$ internal quality factor; $Q_r$ total quality factor; $P_{\text{off}}$ tone power off-resonance; $S_{on}$ median noise power on-resonance; $S_{\text{off}}$ median noise power off-resonance.}
\label{tab:params}
\end{table}

\subsection{1200 GHz KID Array}
\begin{figure}%
    \centering
    \subfloat[\centering]{{\includegraphics[width=6cm]{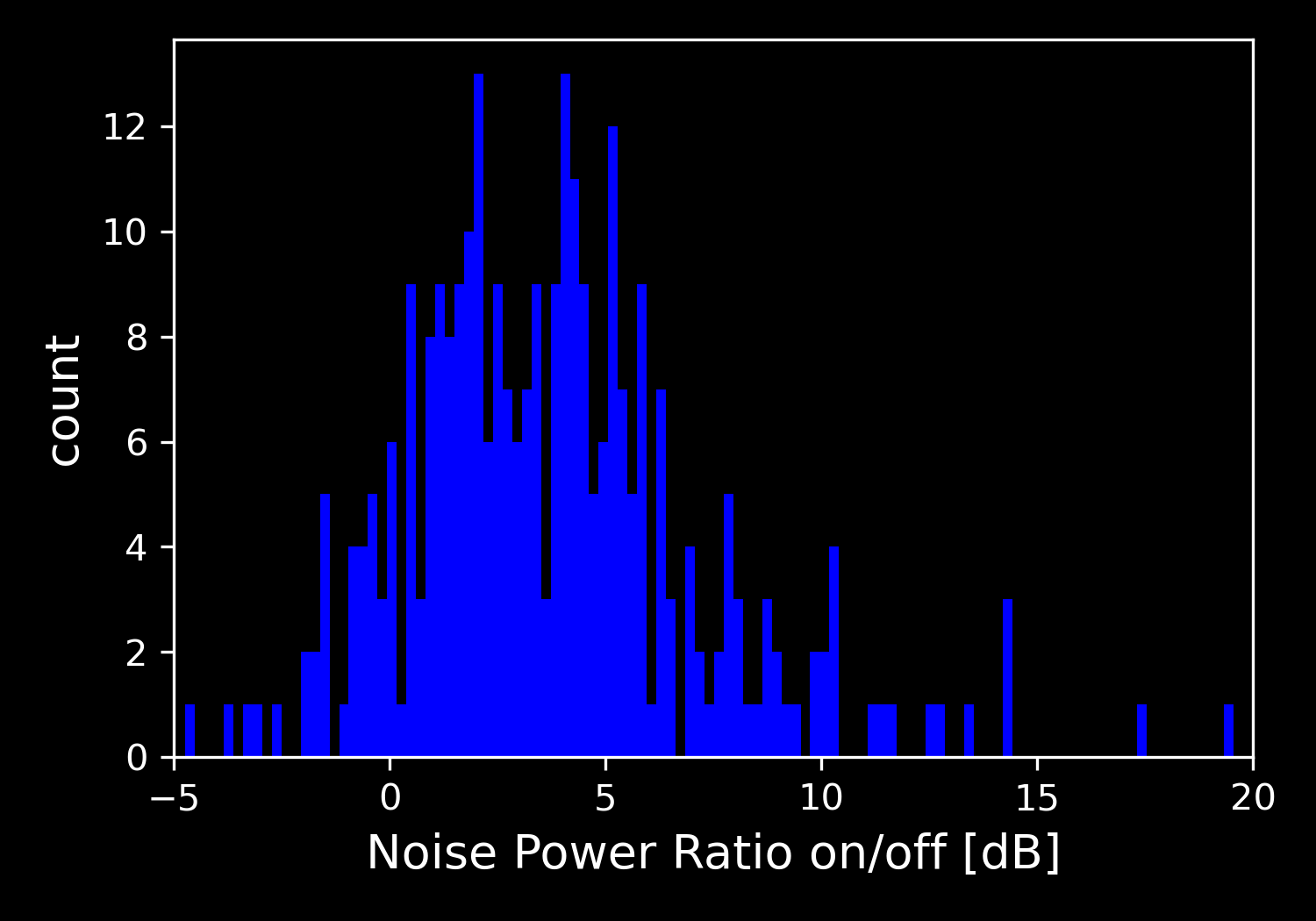} }}%
    \qquad
    \subfloat[\centering]{{\includegraphics[width=6cm]{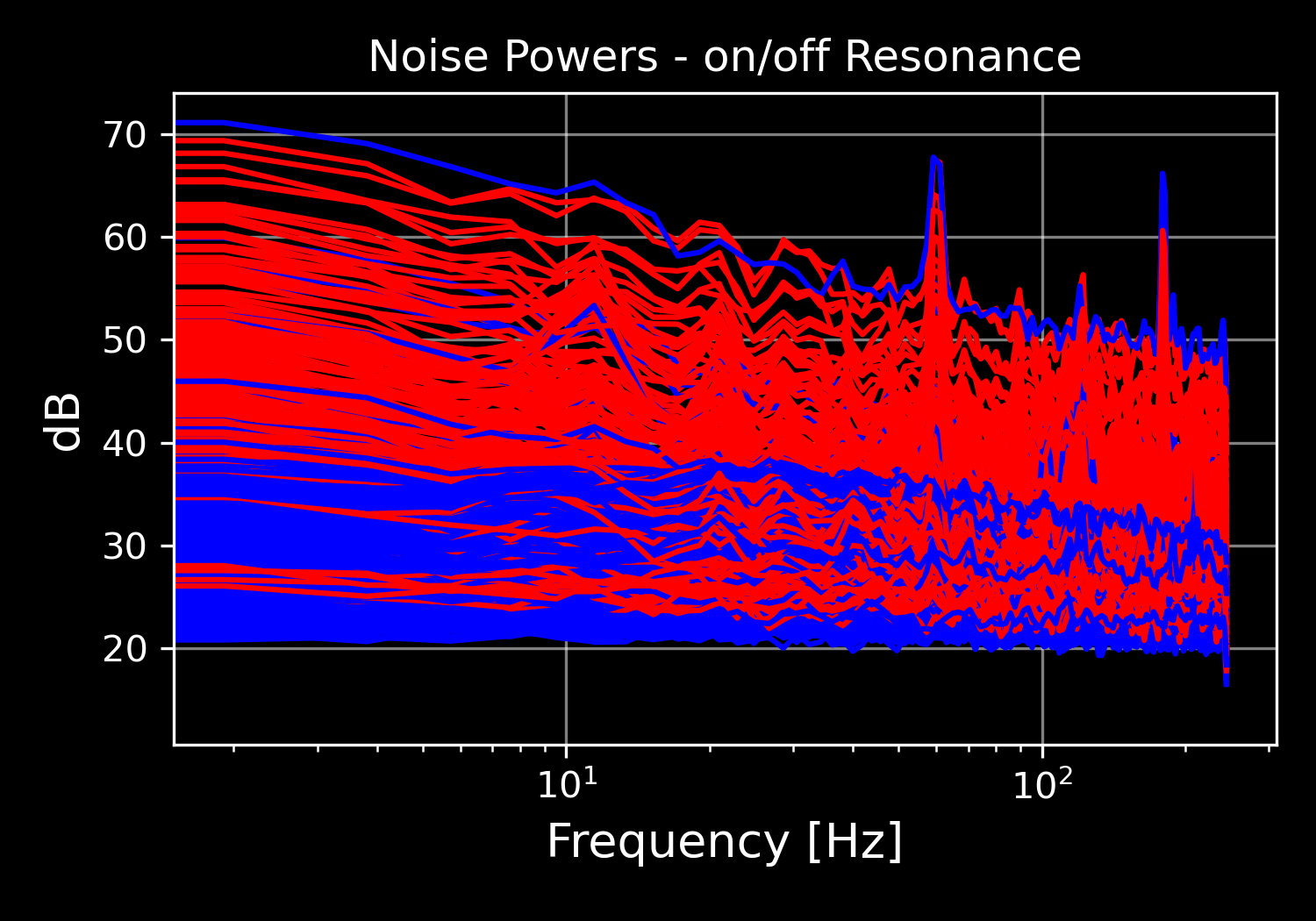} }}%
    \caption{a) Ratio of median noise power spectral density for on and off-resonance for 384 detectors from the 1200 GHz array. b)  Noise power for all 384 detectors from the 1200 GHz array where red is on resonance and blue is off.}%
    \label{fig:1200ghz}%
\end{figure}

Additional measurements were made with a 512 pixel far-infrared (1200 GHz) array intended for an upgraded HAWC+ instrument \cite{Wheeler2022}. These measurements however were solely dark measurements meaning that they did not have the optical stack or source to simulate expected loading conditions. The dark measurements however still provide information on the relationship between the detector and readout noise. Just as with the 280 GHz prototype, we take frequency sweeps along with timestreams on and off-resonance for each detector. Figure \ref{fig:1200ghz} shows a histogram of the noise power ratio on/off-resonance for 384 KIDs from the 1200 GHz array and the noise power spectrum of each detector on the same plot.

\section{Discussion}
The results presented in the previous section show that in some resonators and loading conditions the detector noise power is significantly higher than the combined readout noise power. If this difference between on and off is greater than about 3 dB then we interpret this as detector noise limited performance. It is also observed that in some cases the difference is less than 3 dB and suggests readout noise limited operation. To gain insight into why this is the case we model the effective \textbf{output} noise temperature of the detectors following the methods of Sipola et al. \cite{Sipola2019} and Mauskopf\cite{Mauskopf2018} (solving eq. 107 for T), 
\begin{equation}
    T = \frac{(R_{V} \times \textrm{NEP})^2}{kZ_0}.
\label{eq:T}
\end{equation}
Where NEP is the noise equivalent power, $Z_0$ is the impedance of the system (50 $\Omega$), and k is Boltzmann's constant. $R_{V}$ is the voltage responsivity which can be expanded as, 
\begin{equation}
R_{V}=\frac{dV_{\text{out}}}{dP} = V_{\text{in}}\frac{dx}{dP}\frac{d S_{21}}{dx} \approx V_{\text{in}}\frac{dx}{dP} \frac{2 Q_r^2}{Q_c}.
\end{equation}
The voltage responsivity depends on the input voltage of the bias tone $V_{\text{in}}$, the fractional frequency optical responsivity $dx/dP$, the resonator total quality factor $Q_r$ and coupling quality factor $Q_c$. The last term on the right is a simplification of the derivative of forward transmission applicable when the bias tone is on or very near resonance. This expansion helps to elucidate the detector noise scaling with the various parameters in equation \ref{eq:T},
\begin{equation}
    T \approx \frac{ P_{\text{in}} \textrm{NEP}^2}{k} \Big(\frac{dx}{dP}\Big)^2 \frac{4 Q_r^4}{Q_c^2}.
\label{eq:T_noise}
\end{equation}
Two immediate things to note are that the output detector noise scales linearly with bias tone power, $P_{\text{in}}$, and the intrinsic detector properties scale as the square or higher. From table \ref{tab:params} it can be seen that there is a large variation of noise power between on and off-resonance for different detectors. Using equation \ref{eq:T_noise} we take the measured detector noise $S_{\text{on}} - S_{\text{off}}$ converted to linear units before subtraction and divide this by the measured $P_{\text{in}}$ and $4Q_r^4/Q_c^2$ for each detector. The results are presented in table \ref{tab:discussion}, where the final column gives the resulting term after dividing out the effects of different quality factors and tone powers, leaving the product $NEP^2 (dx/dP)^2$. The uncorrected detector noise powers are within a factor of 9 of each other and the divided values are within a factor of 4 of each other.
This implies that $Q_r^4/Q_c^2$ and $P_{\text{in}}$ can help explain the large differences in detector noise power measured. 

Our analysis demonstrates that optimizing key parameters like input power (Pin), resonator quality factor ratio ($Q_r^4/Q_c^2$), and responsivity slope ($dx/dP$) can significantly reduce readout noise requirements. This translates to a higher effective output noise temperature for the detectors. For illustration, consider Detector 1 data in Table \ref{tab:params}. The on-resonance and off-resonance noise power differ by more than 10 dB. Assuming the Low-Noise Amplifier (LNA) with an input referred noise temperature of 5 K limits off-resonance noise, the effective detector noise temperature must be greater than 50 K. Increasing the input power ($P_{\text{in}}$) by 3 dB could drive the detector noise temperature to 100 K, relaxing the readout noise requirements even further. It should be mentioned that while increasing the bias tone power increases the effective output noise temperature of the detector it simultaneously increases the signal level that we are trying to measure. In other words the signal to noise remains constant if already detector noise limited, or it increases if readout noise limited.
However, higher input power introduces other limitations. Resonator bifurcation can occur at high power levels, where the bias tone dominates dissipation, hindering operation at the resonant frequency. While promising schemes exist to maintain sensitivity under these conditions ( Walker et al.\cite{Walker2024}), they are still under development. Additionally, studies by other groups (Henderson et al.\cite{Henderson2018}, Yu et al.\cite{Yu2022}) highlight non-linear effects in cryogenic amplifiers at higher tone powers.

\begin{table}[]
\centering
\begin{tabular}{llllll}
 Detector - load & $4 Q_r^4/Q_c^2$ & $P_{\text{off}}$ (lin) & $S_{\text{on}}-S_{\text{off}}$ (lin) & NEP$^2(dx/dP)^2$  \\
\hline
1 – 2.9 K   & 2.30E+08  & 2.76E+11  & 1.71E+03   & 2.69E-17 \\
2 – 2.9 K   & 1.79E+08  & 2.19E+11  & 1.03E+03   & 2.63E-17 \\
3 – 2.9 K   & 3.90E+07  & 2.32E+11  & 1.04E+03   & 1.15E-16 \\
4 – 2.9 K   & 3.84E+07  & 1.70E+11  & 1.90E+02   & 2.91E-17 \\
5 – 2.9 K   & 5.57E+07  & 1.45E+11  & 1.90E+02   & 2.36E-17 
\end{tabular}
\caption{280 GHz array with the $Q_r^4/Q_c^2$ scaling term, the tone power, detector noise power, and the NEP$^2(dx/dP)^2$ term which is the detector power divided by the previous two columns.}
\label{tab:discussion}
\end{table}

\section{conclusion}
We have presented preliminary results showing detector noise limited performance while utilizing the RFSoC-based readout on multiple KID arrays.
An output noise temperature viewpoint was taken to help explain the differences in the measured detector noise powers and also highlight the important terms to maximize for relaxing readout noise requirements. We expect the science grade arrays for Mod-Cam and Prime-Cam to have properties similar to detectors 1 and 2 from table \ref{tab:params} and thus assure not only detector noise limited readout operation but also suggest the possibility of background limited on-sky operation.

\appendix    
\section{Derivation of T(x)/T(0)}
Equation \ref{eq:T} for the effective output noise temperature depends on the bias tone frequency placement. We denote the fractional frequency shift from resonance as $x=f/f_0 - 1$. To find the change in the output noise temperature with shifts away from resonance, we normalize $T(x)$ with its value on resonance $T(x=0)$,
\begin{equation}
    \frac{T(x)}{T(0)} = \Big[\Big( \frac{d Re\{ S_{21}(x)\}}{dx}\Big)^2 + \Big( \frac{d Im\{ S_{21}(x)\}}{dx} \Big)^2\Big] \Bigg/ \Big( \frac{d Im\{ S_{21}(x)\}}{dx}\Big)^2\Bigg|_{x=0},
\end{equation}
where $S_{21}(x)$ is the complex forward transmission of the resonator. We sum the noise power of both the real and imaginary components. Expanding the forward transmission derivatives we get,
\begin{equation}
    = \Big[\Big(\frac{8 Q_r^{3} x}{Q_c (1 + 4 Q_r^{2} x^2)^{2}}\Big)^2 + \Big(\frac{2 Q_r^{2}}{Q_c (1 + 4 Q_r^{2} x^2)} - \frac{16 Q_r^{4}x^2}{Q_c (1 + 4 Q_r^{2} x^2)^{2}}\Big)^2\Big] \Bigg/ \frac{4 Q_r^4}{Q_c^2},
\end{equation}
which reduces to,
\begin{equation}
 = \frac{4 Q_r^4}{Q_c^2 (1 + 4 Q_r^2 x^2)} \Big/ \frac{4 Q_r^4}{Q_c^2} = \frac{1}{(1 + 4 Q_r^2 x^2)^2} = \frac{1}{(1 + 4 N^2)^2},
\end{equation}
replacing $x Q_r$ with N, the number of linewidths.

\acknowledgments %
The CCAT-prime project, FYST and Prime-Cam instrument have been supported by generous contributions from the Fred M. Young, Jr. Charitable Trust, Cornell University, and the Canada Foundation for Innovation and the Provinces of Ontario, Alberta, and British Columbia. The construction of the FYST telescope was supported by the Gro{\ss}ger{\"a}te-Programm of the German Science Foundation (Deutsche Forschungsgemeinschaft, DFG) under grant INST 216/733-1 FUGG, as well as funding from Universit{\"a}t zu K{\"o}ln, Universit{\"a}t Bonn and the Max Planck Institut f{\"u}r Astrophysik, Garching.
The construction of EoR-Spec is supported by NSF grant AST-2009767. The construction of the 350 GHz instrument module for Prime-Cam is supported by NSF grant AST-2117631.

The author would also like to thank the NRC Herzberg mm-wave lab for use of some measurement equipment and Logan Foote for evaluating our RFSoC-based system and bringing readout bugs to our attention.

\bibliography{main} 
\bibliographystyle{spiebib} 

\end{document}